\documentclass[twocolumn,showpacs,amsmath,amssymb,aps,reprint]{revtex4-1}
\usepackage[breaklinks=true,colorlinks=true,linkcolor=blue,urlcolor=blue,citecolor=blue]{hyperref}
\usepackage{graphicx}%
\usepackage{graphics}%
\usepackage{soul,color}
\usepackage{epstopdf}
\usepackage{dcolumn}%
\usepackage{bm}% bold math

\begin{document}

\title{Tunable magnetism on the lateral mesoscale\\ by post-processing of Co/Pt heterostructures}
\author{Oleksandr V. Dobrovolskiy$^{1,2*}$, Maksym Kompaniiets$^1$, Roland Sachser$^1$, Fabrizio Porrati$^1$, Christian Gspan$^3$, Harald Plank$^{3,4}$, Michael Huth$^1$}
\address
        {
        $^1$Physikalisches Institut, Goethe University, 60438 Frankfurt am Main, Germany\\
        $^2$Physics Department, V. Karazin National University, 61077 Kharkiv, Ukraine\\
        $^3$Graz Centre for Electron Microscopy, 8010 Graz, Austria\\
        $^4$Institute for Electron Microscopy and Nanoanalysis, TU Graz, 8010 Graz, Austria
        }

\begin{abstract}
Controlling magnetic properties on the nm-scale is essential for basic research in micro-magnetism and spin-dependent transport, as well as for various applications such as magnetic recording, imaging and sensing. This has been accomplished to a very high degree by means of layered heterostructures in the vertical dimension. Here we present a complementary approach that allows for a controlled tuning of the magnetic properties of Co/Pt heterostructures on the lateral mesoscale. By means of in-situ post-processing of Pt- and Co-based nano-stripes prepared by focused electron beam induced deposition (FEBID) we are able to locally tune their coercive field and remanent magnetization. Whereas single Co-FEBID nano-stripes show no hysteresis, we find hard-magnetic behavior for post-processed Co/Pt nano-stripes with coercive fields up to 850 Oe. We attribute the observed effects to the locally controlled formation of the CoPt L1$_{0}$ phase, whose presence has been revealed by transmission electron microscopy.
\end{abstract}

%\keywords{focused electron beam induced deposition; in-situ processing; heterostructures; cobalt; platinum}

\maketitle
%\tableofcontents

\section{Introduction}
Controlling magneto-transport properties on the nm-scale is essential for basic research in micro-magnetism~\cite{Kro03boo} and spin-dependent transport~\cite{Kei06nat} as well as for various applications, such as magnetic domain-wall  logic~\cite{All05sci} and memory~\cite{Par08sci}, fabrication of Hall sensors~\cite{Gab10nan} and cantilever tips~\cite{Bel12rsi} for magnetic force microscopy (MFM). In particular, the ability to tune the magnetization is the basic property needed for the realization of stacked nanomagnets~\cite{Tak06jap}, pinning of magnetic domain walls~\cite{Bri11prl} and Abrikosov vortices~\cite{Vel08mmm,Dob10sst,Dob11pcs}, magnetic sensing~\cite{Gab10nan,Fer09apl} and storage~\cite{Par08sci,All05sci}, and spin-triplet proximity-induced superconductivity~\cite{Buz05rmp,Ber05rmp,Wan10nat,Kom14apl,Kom14jap}. This magnetization tuning has been accomplished to a very high degree by means of layered heterostructures in the vertical dimension, which can be prepared by thin film techniques or by an alternative approach -- as used by us -- by direct writing of metal-based layers by focused electron beam induced deposition (FEBID)~\cite{Utk08vst,Hut12bjn}. The resolution of FEBID is better than $10$~nm laterally and $1$~nm vertically~\cite{Utk08vst,Hut12bjn} and, thus, its proven applications range from photomask repair~\cite{Lia05vst} to fabrication of nanowires~\cite{Fer13scr,Kom14jap}, nanopores~\cite{Dan06lan}, magnetic~\cite{Gab10nan,Fer09apl} and strain sensors~\cite{Sch10sen} as well as direct-write superconductors~\cite{Win14apl}.~
\begin{figure*}
    \includegraphics[width=0.85\textwidth]{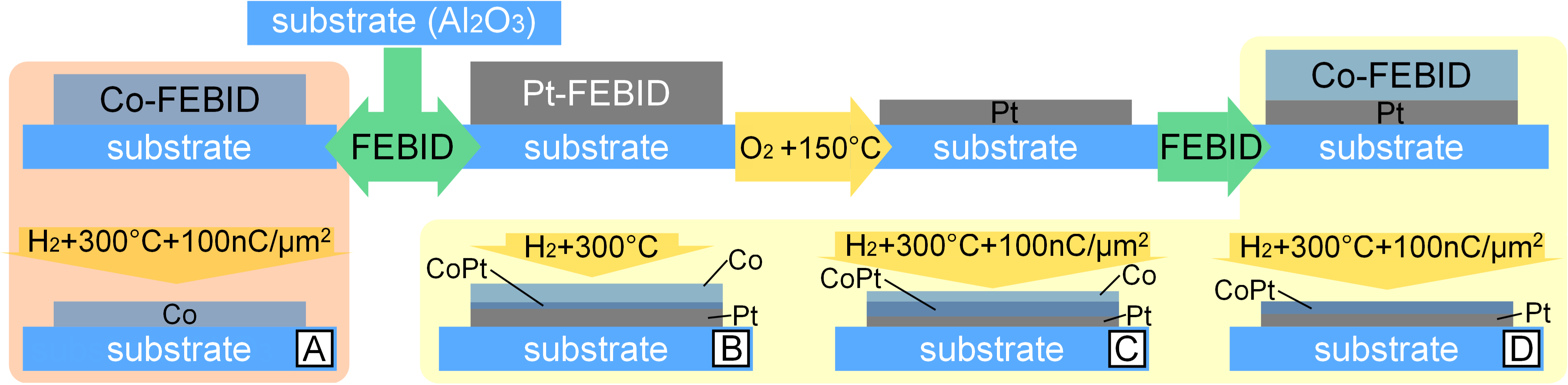}
     \caption{Preparation and post-processing of the samples investigated in this work. Throughout the text the samples are referred to by their labels A, B, C, and D, as indicated.}
    \label{fSketch}
\end{figure*}
\begin{figure*}
    \includegraphics[width=0.75\textwidth]{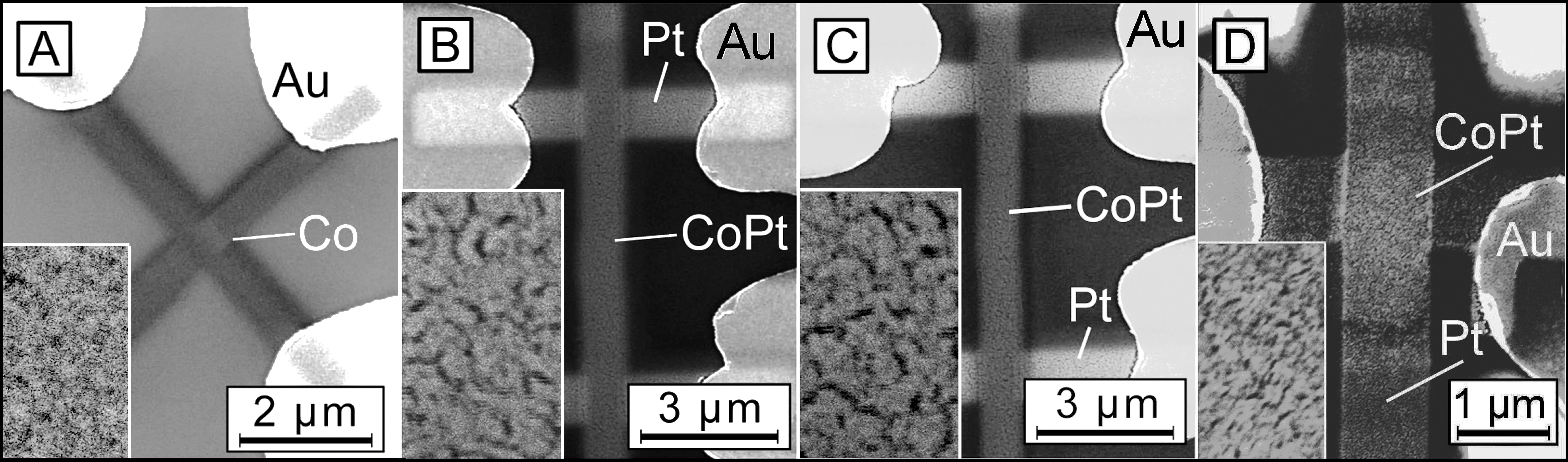}
    \caption{SEM images of the samples. The $500\times860$~nm$^2$ insets show the morphology of the post-processed Co/Pt FEBID nano-stripes in the middle of the overlap of the nano-stripes.}
    \label{fSEM}
\end{figure*}

The precursors Co$_2$(CO)$_8$ and (CH$_3$)$_3$CH$_3$PtC$_5$H$_4$ from which Pt- an Co-based structures can be fabricated in the FEBID process, like most metal-organic precursors, \emph{do not dissociate into the respective pure materials}, unless FEBID of Co is done at elevated substrate temperatures~\cite{Bel11nan}. By contrast, when decomposed in the focus of the electron beam into volatile components and the permanent deposit on the processed surface, these form \emph{granular} metals, whose grains are embedded in a carbon-rich, poorly conducting matrix. In consequence, the electrical conductivity of as-deposited Pt-based FEBID structures usually is in the high-ohmic or even the insulating regime while that of as-deposited Co-FEBID structures is at least an order of magnitude lower than that of pure Co, typically. In addition, though the magnetic properties of as-deposited Co-FEBID structures are sufficient for application in Co MFM tips~\cite{Bel12rsi} and studying the effects of topological structures on the magnetization reversal process~\cite{Lar14apl}, these properties differ from those of pure Co. Still, owing to the sensitivity of the matrix to post-processing treatments, the compositional, structural, and, hence, electrical~\cite{Sac11prl,Por11jap} and magnetic~\cite{Ber10cie,Beg15nan} properties of metal-based layers fabricated by FEBID can be substantially modified either in-situ or ex-situ. Exemplary purification treatments of samples include annealing in reactive gases~\cite{Bot09nan}, electron irradiation~\cite{Sac11prl,Por11jap}, or a combination of both~\cite{Meh13nan,Pla14ami,Gei14jpc,Beg15nan}.~

Several approaches have already been proposed for the preparation of magnetic nanoparticles and their alloying, in particular, with the purpose of eventual using them for ultrahigh-density data-storage media. Thus, driven by the need to accomplish the above demand, FePt magnetic nanoparticles were prepared using colloidal chemistry~\cite{Sun00sci} and micellar methods~\cite{Eth07adm}. The latter method was also extended to the preparation of CoPt nanoparticles~\cite{Wie10bjn}. Later on, it turned out easier to deposit self-assembled Co nanoparticles on top of Pt thin films~\cite{Han11bjn} and thereby fabricate surface alloys formed at step edges of Pt single crystalline substrates. In that work~\cite{Han11bjn}, an increase of the coercive field and of the Co orbital magnetic momentum was observed and attributed to the formation of the CoPt L1$_0$ phase with strongly increased magnetic anisotropy compared to pure Co.~

Here, we employ direct writing of Pt and Co layers by FEBID and demonstrate by means of in-situ post-processing how to \emph{locally} tune the coercive field and the remanent magnetization of layered Co/Pt FEBID nano-stripes. This is achieved by a combination of in-situ heating in a local reactive gas atmosphere (H$_2$ and O$_2$) and electron-beam irradiation of as-deposited layers, as is sketched in Fig.~\ref{fSketch}. We show that the magnetic response of the nano-stripes can be tuned on the lateral mesoscale, from the magnetic properties of Co to the hard ferromagnetic response of the CoPt L1$_0$ phase, whose presence has been revealed by transmission electron microscopy.~

\section{Experimental methods}
\subsection{Preparations and geometry}
Co and Pt growth, processing and imaging experiments were carried out in a dual-beam high-resolution scanning electron microscope~(SEM: FEI, Nova NanoLab 600). The SEM was equip\-ped with a multi-channel gas injection system for FEBID. As substrates we used epi-polished c-cut (0001) Al$_2$O$_3$ with Cr/Au contacts of 3/50 nm thickness prepared by photolithography in conjunction with lift-off. The samples are one Co-FEBID structure and three Co/Pt-FEBID nano-stripes labeled as sample A, B, C, and D, respectively. The Co/Pt deposits B and C bridging a $12~\mu$m gap between the Au contacts were deposited in a 6-point geometry, while samples A and D were deposited in a cross-shaped fashion, see Fig.~\ref{fSEM} for an overview. The only reason for the different geometry of samples B and C lies in that they were originally designed for other measurements in addition to those reported here.~

\subsection{FEBID of Pt}
\begin{figure}
    \centering
        \includegraphics[width=0.9\linewidth]{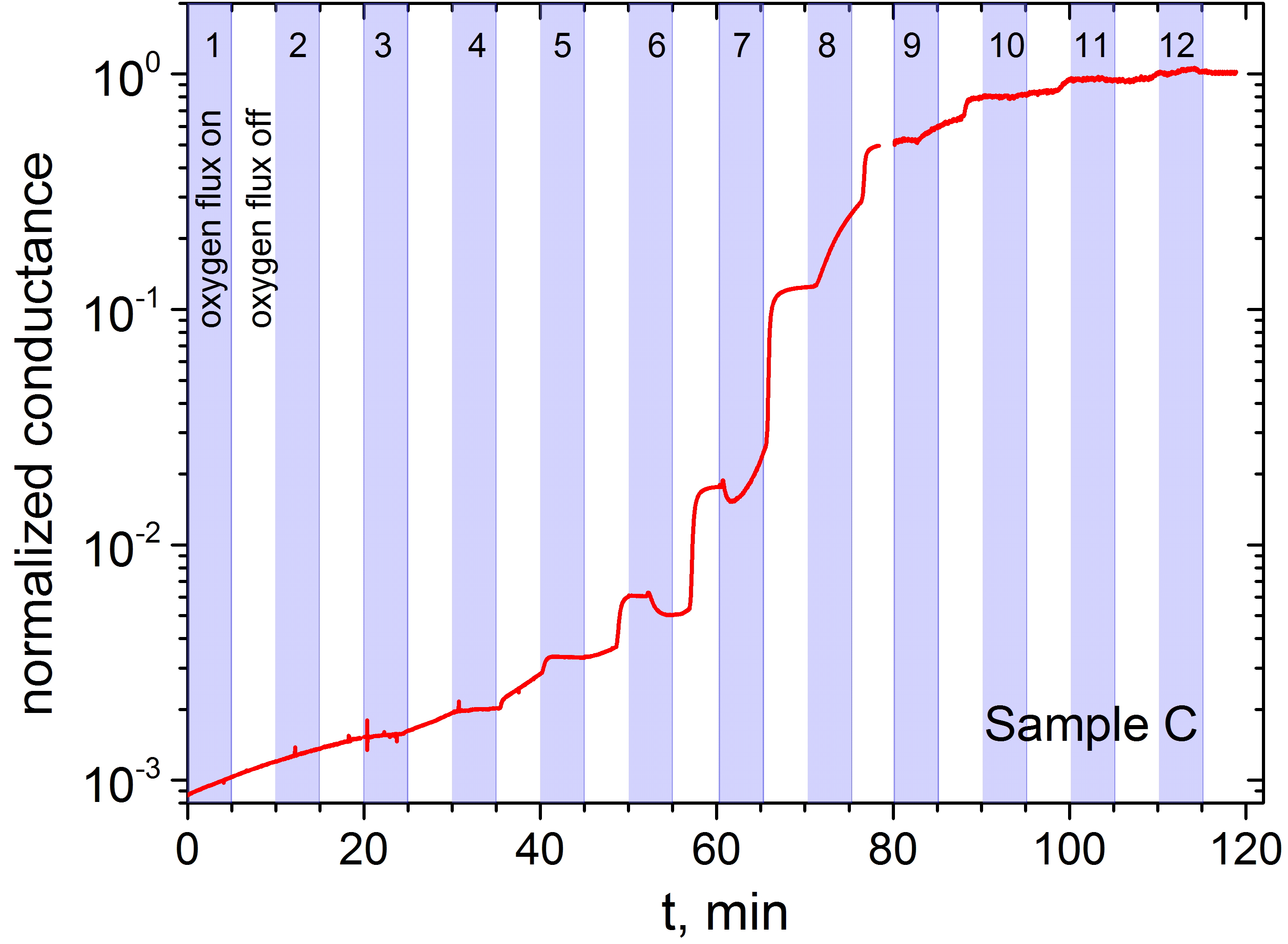}
        \caption{Time-dependent conductance of the Pt layer of sample C normalized to its saturation value after the purification process at $150^\circ$C consisting of 12 cycles with a total duration of 120 min. The filled areas under the curve show the time intervals with the oxygen flux switched on.}
    \label{fInSitu}
\end{figure}

FEBID of Pt was used for the fabrication of the bottom layers of all samples, with exception of sample A. In the FEBID process the precursor gas was (CH$_3$)$_3$Pt(CpCH$_3$), the beam parameters were $5$~kV/$1$~nA, the pitch was $20$~nm, the dwell time was $1~\mu$s, the precursor temperature was $44^\circ$C, and the process pressure was $9.5 \times 10^{-6}$~mbar for a needle position of the gas injector at $100~\mu$m height and $100~\mu$m lateral shift from the writing field position. After the Pt deposition, the samples were heated up to $150^\circ$C in the same SEM without breaking the vacuum. For the design of the heatable stage adapter and the sample holder we refer to Ref.~\cite{Sac14ami}. Once heated, the Pt-based deposits were subjected to an oxygen flux fed into the vacuum chamber up to a pressure of $1.5\times10^{-5}$~mbar through a home-made gas injection system. The samples were subjected to 12 cycles of oxygen flux switched on for 5 minutes interrupted by 5-minute turn-offs. The resistivity of the as-deposited Pt-based layers was $0.4~\Omega$cm, decreased to about  $90$\,m$\Omega$cm as the temperature rose to $150^\circ$C, and dropped to $70$-$90~\mu\Omega$cm after 10 oxygen pulses. Figure~\ref{fInSitu} depicts the time-dependent normalized conductance of the Pt layer of sample C during the purification process. The post-processed Pt layers exhibited a nano-porous structure and a reduction of height from $50\pm1.5$~nm to $11\pm1.5$~nm, as inferred from atomic force microscopy, due to the removal of the carbonaceous matrix~\cite{Sac14ami}. The void volume fraction of the very thin purified Pt layer was estimated from a grey scale threshold analysis of the SEM image which yields a value of $0.31\pm0.07$.~

\subsection{FEBID of Co}
FEBID of Co was used for the preparation of the top layers of the structures. In the FEBID process the precursor gas was Co$_2$(CO)$_8$, the beam parameters were $5$~kV/$1$~nA, the pitch was $20$~nm, the dwell time was $50~\mu$s, the precursor temperature was $27^\circ$C, and the process pressure was $8.85 \times 	10^{-6}$~mbar. Before the deposition, the chamber was evacuated down 	to $4.12\times10^{-6}$~mbar. For removing the residual water from the SEM chamber a home-made liquid-nitrogen trap filled with zeolite powder was employed. After the deposition all samples were heated up to $300^\circ$C in the SEM without breaking the vacuum and subjected to a H$_2$ flux fed into the SEM chamber up to a pressure of $1.5\times10^{-5}$~mbar. While kept at $300^\circ$C, samples A and C, and D were additionally irradiated with the electron beam ($5$~kV/$1$~nA, $20$~nm pitch, $50~\mu$s dwell time), whereas sample B was left non-irradiated. The irradiation dose was $100$~nC/$\mu$m$^2$ for all irradiated samples. After this purification step, the thickness of the Co layers reduced by a factor of 1.55, in agreement with previous work~\cite{Beg15nan}.~

\subsection{Thickness-integrated EDX}
The thickness-integrated material composition of the samples was inferred from energy-dispersive X-ray (EDX) spectroscopy, in the same SEM without exposure of the deposits to air. The EDX parameters were $5$~kV and $1.6$~nA. The elemental composition was calculated taking into account ZAF (atomic number, absorbtion and fluorescence) and background corrections. The software we used to analyze the material composition in the deposits was EDAX's Genesis Spectrum v. 5.11. The elemental composition was quantified without thickness correction, so that the reported data are a qualitative indicator only.~

\subsection{Electrical resistance measurements}
The electrical and magneto-resistance measurements were carried out in a helium-flow cryostat equipped with a superconducting solenoid. The measurements were done in the current-drive mode, with a current density of the order of $10$\,kA/cm$^2$. For the Hall voltage measurements a lock-in amplifier in conjunction with a differential preamplifier and a ratio transformer to null the signal at $H = 0$ were used~\cite{Bey05ajp}. The measurements were done with the magnetic field directed normally to the stripe plane and immediately after transferring the samples from the SEM after the Co purification step.~

\subsection{Transmission electron microscopy}
For an inspection of the selected sample C by scanning transmission electron microscopy (STEM) a Titan G2 microscope from FEI with a CS probe corrector (DCOR) was used. The TEM was equipped with a X-FEG high-brightness electron gun, the high-end post-column electron energy filter Quantum ERSTM from Gatan, and four high sensitivity SDD X-ray detectors from Bruker (Super-X). The measurements were performed at an accelerating voltage of $300$~kV with an electron probe diameter smaller than $1$~\AA. Before the TEM measurements, sample C was covered with a $300$~nm-thick protective Pt-C layer deposited by FEBID. The pixel time for the energy-dispersive x-ray cross-sectional line scan (cross-sectional EDX) was 8 seconds per spectrum and the step size was 0.8 nm. ~

\subsection{Nano-diffraction and simulations}
A convergence angle of $1.0$ mrad was used to generate electron nanodiffraction patterns in the STEM mode. These diffraction patterns were recorded energy-filtered on a 16-bit CCD. To collect the nanodiffraction images over the complete layer the ``diffraction spectrum image'' technique was used as part of the software package Digital Micrograph (Gatan). The lateral step size from pixel to pixel was $3.7$~nm. Therefore, an individual selection of the diffraction patterns from the upper and the lower layer was possible. For a comparison with the experimental nanodiffraction data from the upper and lower layer, electron diffraction simulations for the CoPt fcc- and fct-phase assuming bulk lattice constants were made with the software JEMS~\cite{Sta12ems}. The simulations were done in the kinematic mode. For the generation of the elemental signal profile the intensity from the Pt M edge (2.05 keV) and the Co K edge (6.92 keV) was used. ~

\section{Results and Discussion}
\subsection{Structural and electrical resistance properties}
SEM images of the samples investigated in this work are shown in Fig.~\ref{fSEM}, while their geometrical dimensions, elemental composition, and magnetic properties are compiled in Table~\ref{Table}.~
\begin{table}[h!]
\begin{tabular}{|l|l|l|l|l|l|l|l|l|l|l|l|}\hline
\bfseries S.   & $l$,      & $w$,     & $d_{\mathrm{Co}}$, & $d_{\mathrm{Pt}}$,& Co,        & Pt,       & C,         & $H_c$, & $H_s$ & $M_r/$ & Co/      \\
\bfseries \#         & $\mu$m & $\mu$m & nm                        & nm                       & at.$\%$ & at.$\%$ & at.$\%$ &  Oe     & T       &  $M_s$        & Pt            \\\hline
A                       & 0.49     & 0.5       & 11                         & 0                        & 92        & 0          & 8         &$\times$& 1.7     &$\times$    & $\infty$   \\\hline
B                       & 5.45     & 1          & 10                        & 11                       & 54        & 27      & 19     & 770     & 1.5     & 0.15        &  2           \\\hline
C                       & 5.35     & 1          & 11                        & 11                       & 49        & 22        & 29       & 850     & 1.3     & 0.25         & 2.23       \\\hline
D                       & 1         & 1          & 5                         & 11                        & 35       & 35         & 30       & 420     & 0.5     & 0.18         &1            \\\hline
\end{tabular}
\caption{Geometrical dimensions, thickness-integrated composition, and magnetic properties of the samples. $l$:~length; $w$: width; $d_{\mathrm{Co}}$: thickness of the Co layer; $d_{\mathrm{Pt}}$: thickness of the Pt layer; $H_c$: coercive field; $H_s$: saturation field; $M_r/M_s$: remanent-to-saturation magnetization ratio (squareness).}
\label{Table}
\end{table}
\begin{figure*}
    \includegraphics[width=1\textwidth]{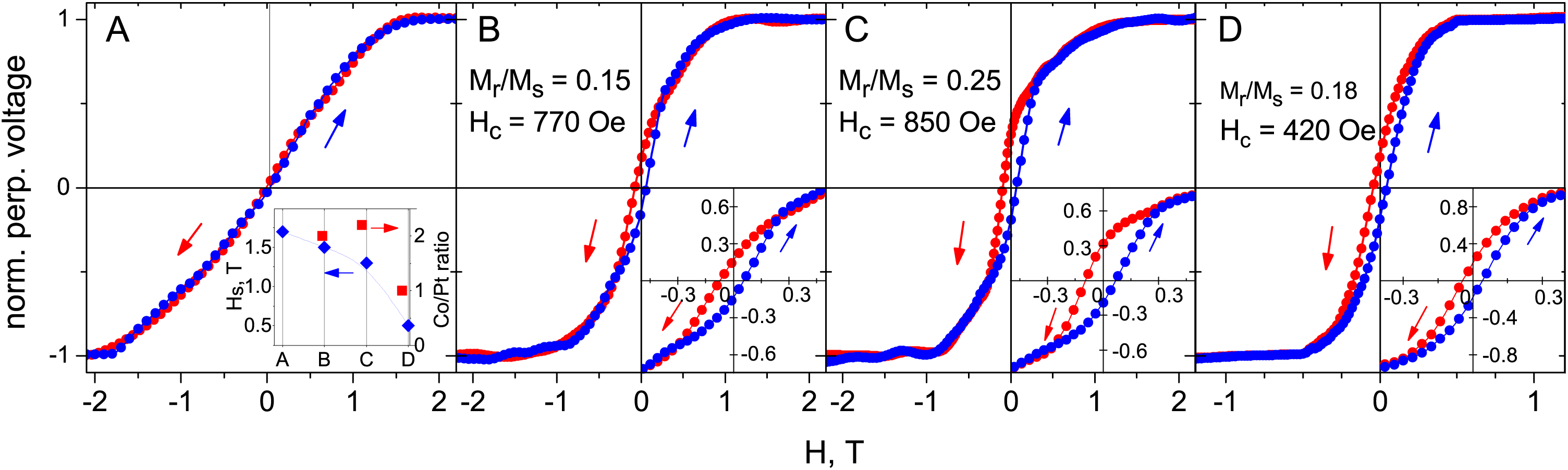}
    \caption{Hall voltage cycling at 10~K for all samples. Before measurements, all samples were saturated at 3~T. Note the different field range and scale for sample D. Inset in A: The magnetization saturation fields $H_s$ and the Co/Pt ratios for all samples.}
    \label{fHall}
\end{figure*}

The EDX data were acquired at the overlaps of the nano-stripes and normalized to $100$\, at.$\%$ after exclusion of the oxygen-based signal whose bulk part unavoidably stems from the substrate (Al$_2$O$_3$), due to the small thickness of the investigated samples. At the same time, from previous work~\cite{Beg15nan} where we reported, in particular, a reduction of the oxygen content in individual Co stripes at different stages of the same purification treatment, we aware of the remaining O content at a level of about $10$\,at.$\%$ in the processed stripes. For these reasons, though acquired with a statistical error of $3\%$, the EDX data in table~\ref{Table} only serve as an indicator of the Co/Pt ratio being crucial for the different Co/Pt alloy phase formation --- an issue to which we return in what follows.~

The temperature dependence of the electrical resistance of all samples is metallic. The resistivities of the samples at 10~K are about $40~\mu\Omega$cm and the room temperature-to-10~K resistance ratios are about 1.3. The room temperature resistivity values are an order magnitude larger than the literature values for bulk Co and Pt~\cite{Kit04boo} and are in agreement with the recently reported values for purified individual Co~\cite{Beg15nan} and Pt~\cite{Sac14ami} FEBID structures.~

\subsection{Magneto-transport properties}
The central finding of this work lies in the modification of the field dependences of the Hall voltage $U(H)$ measured at 10~K for all samples, see Fig.~\ref{fHall}. The magnetic field was directed perpendicular to the sample plane and, hence, the out-of-plane magnetization was probed by the measurements. This means that first the shape anisotropy of the stripe had to be overcome and all recorded loops relate to the hard-axis magnetization behavior.~

The reference Co-based sample A shows no hysteresis, whereby $U(H)$ is nearly linear from $-1.5$~T to $1.5$~T and saturates at $H_s = \pm1.7$~T. The $U(H)$ curve of the Co/Pt-based sample B demonstrates two distinctive features compared to sample A: Sample B shows a noticeable hysteresis loop and its saturation field $H_s$ is by about $30\%$ smaller than $H_s$ for sample A. The behavior of sample B is that of ferromagnet, with a coercive field $H_c$ of $770$~Oe and a remanent-to-saturation magnetization ratio (squareness) $M_r/M_s$ of $0.15$. The irradiated Co/Pt-based sample C exhibits an even broader hysteresis loop with $H_c = 850$~Oe and $M_r/M_s = 0.25$, respectively, and its saturation field $H_s$ amounts to $1.3$~T. Even though samples B and C demonstrate a hysteresis loop, we note that it is not completely open and the overall behavior of the Hall voltage curves is suggestive of a superposition of a soft and hard ferromagnetic response. We attribute these contributions to different phases formed at different depths within the layered nano-stripe, as will be corroborated by a TEM inspection in the section devoted to the microstructure analysis.~

Summarizing this part, the following two effects are observed in the post-processed Co/Pt samples, namely (i) a \emph{hysteresis development} and (ii)~a \emph{reduction of the saturation field}. To explain both effects, we next discuss the processes which take place in the deposits in the course of purification treatments.~

\subsection{Purification mechanisms}
The as-deposited reference sample A has a nanogranular Co microstructure with inclusions of carbon and oxygen. The employed purification procedure of heating at $300^\circ$C in H$_2$ atmosphere in conjunction with electron irradiation relies upon the Fischer-Tropsch reaction~\cite{And84boo,Beg15nan}. In this chemical process, cobalt serves as a catalyst, while volatile hydrocarbons and water are produced, effectively oxidizing the carbon. Thus, in the course of the reaction, carbon is partially removed from the deposit causing a reduction of the deposit thickness. The magnetic behavior of the thin polycrystalline Co stripe A is dominated not by the magnetocrystalline anisotropy, but rather by the shape anisotropy causing the magnetization to lie preferentially along the stripe axis. Given the demagnetizing factor for the created geometry, $N\approx 1$~\cite{Hub08boo}, we arrive at a saturation magnetization of $M_s = H_s/N = 1.7~\mathrm{T}\times10^4/4\pi \cong 1353$~emu/cm$^3$, corresponding to $98\,\%$ of the bulk value~\cite{Cha98sci}. Allowing for an up to $5\%$ error in the determination of the saturation magnetization value and a concurrent of the presence of carbon and oxygen in sample A, this value is likely slightly overestimated and, hence, should be regarded as an upper bound only.~

The as-deposited Pt-FEBID layers for samples B and C are also nanogranular metals. The purification mechanism for Pt-FEBID structures relies upon the catalytic activity of Pt~\cite{Win97nat,Sac14ami} in oxygen atmosphere. Namely, when delivered close to the deposit surface, molecular oxygen is dissociatively chemisorbed on the surface of the metallic Pt particles. Since the process takes place at $150^{\circ}$C, a thermally activated oxidation of carbon at the Pt/C interface occurs, leading to the formation of CO and a reorganization and coalescence of Pt nanocrystallites by surface diffusion. The latter, in turn, results in a nanoporous morphology which is clearly seen in the SEM images of samples B and C in the insets to Fig.~\ref{fSEM}. As will be shown below by TEM it is this nanoporosity which allows Co to penetrate into the Pt layer during the Co deposition and to form a Co/Pt alloy phase. Considering the Co-Pt binary phase diagram~\cite{Mas90boo}, for a Co/Pt-ratio of $1$:$1$, the CoPt L1$_0$ phase can form. This phase is a hard ferromagnet whose presence can explain both, the reduction of the saturation field as well as the appearance of a hysteresis loop in samples B and C.
\subsection{Microstructure analysis}
\begin{figure}
    \includegraphics[width=0.6\linewidth]{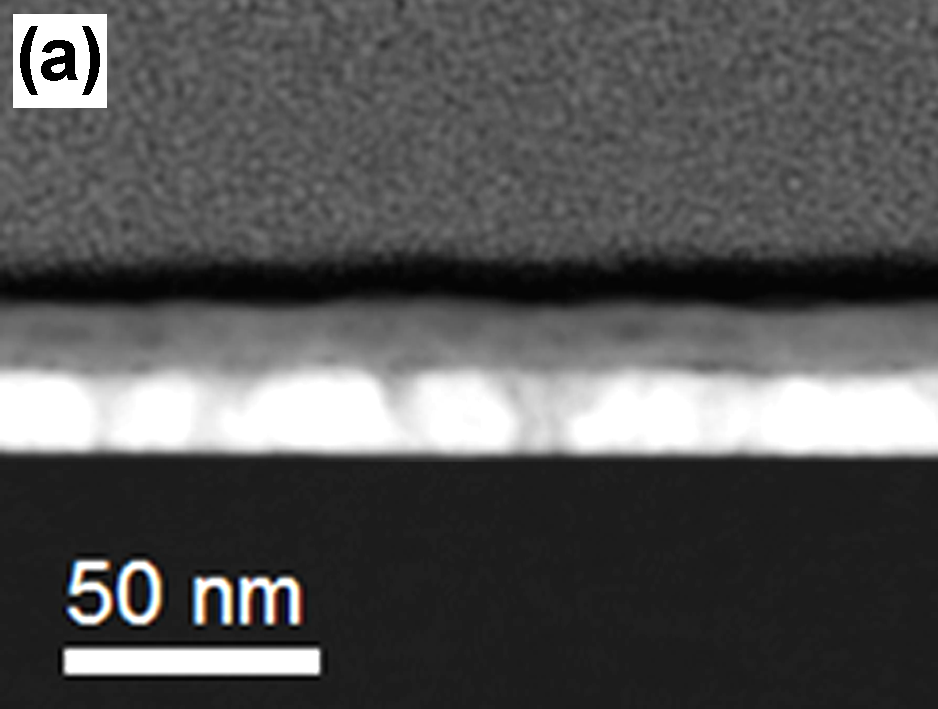}\\
    \hspace{0.8cm}
    \includegraphics[width=0.7\linewidth]{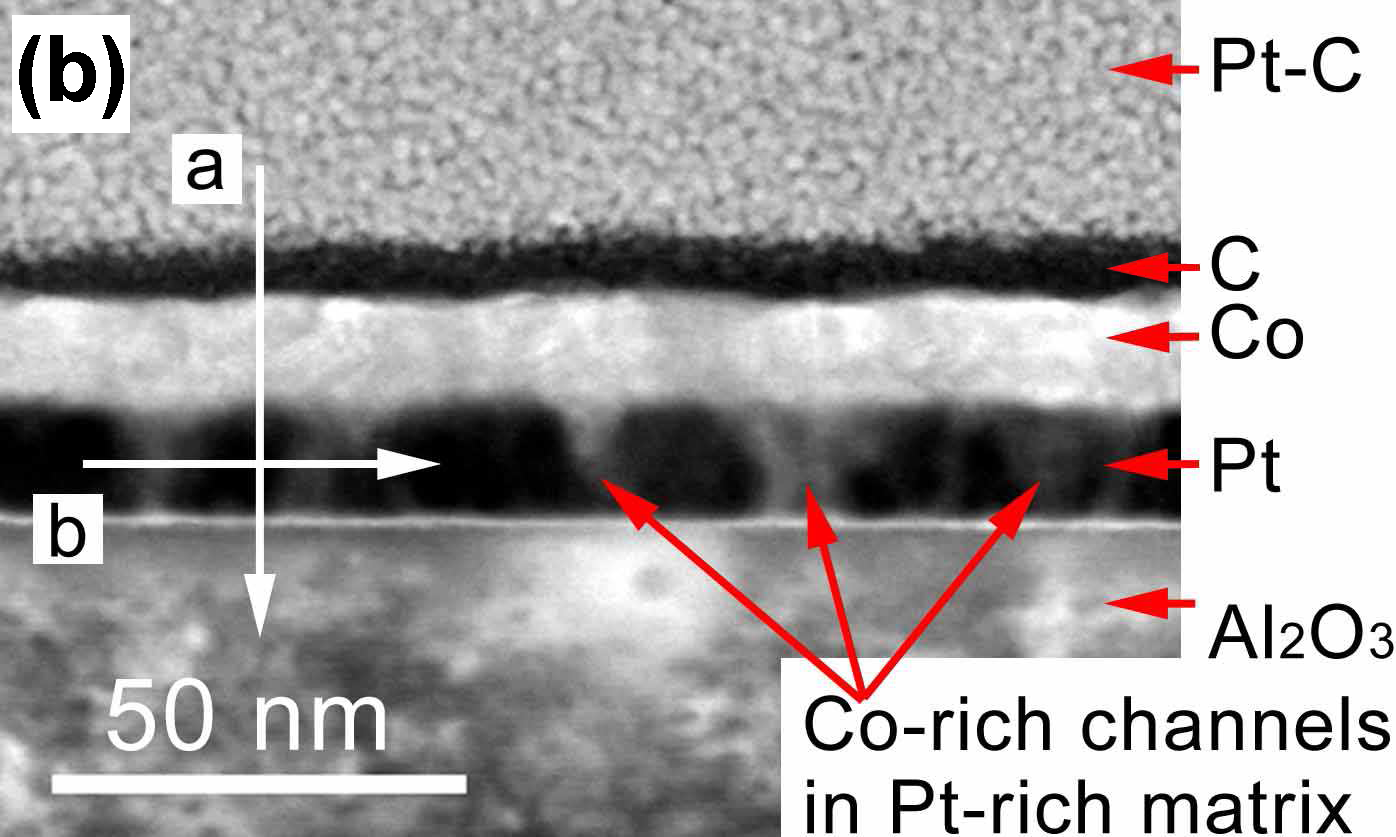}
    \caption{TEM micrographs of sample C acquired (a) in the high angle annular dark field mode and (b) in the annular dark field mode. In (a), elements with higher atomic numbers $Z$ are brighter in the image. The light regions in the Pt layer in (b) correspond to Co-rich channels embedded in the Pt-rich matrix. The arrows depict the directions along which the STEM-EDX elemental peak intensites in Fig.~\ref{fEDX} have been acquired.}
   \label{fADF}
\end{figure}

\begin{figure}
\centering
    \includegraphics[width=0.7\linewidth]{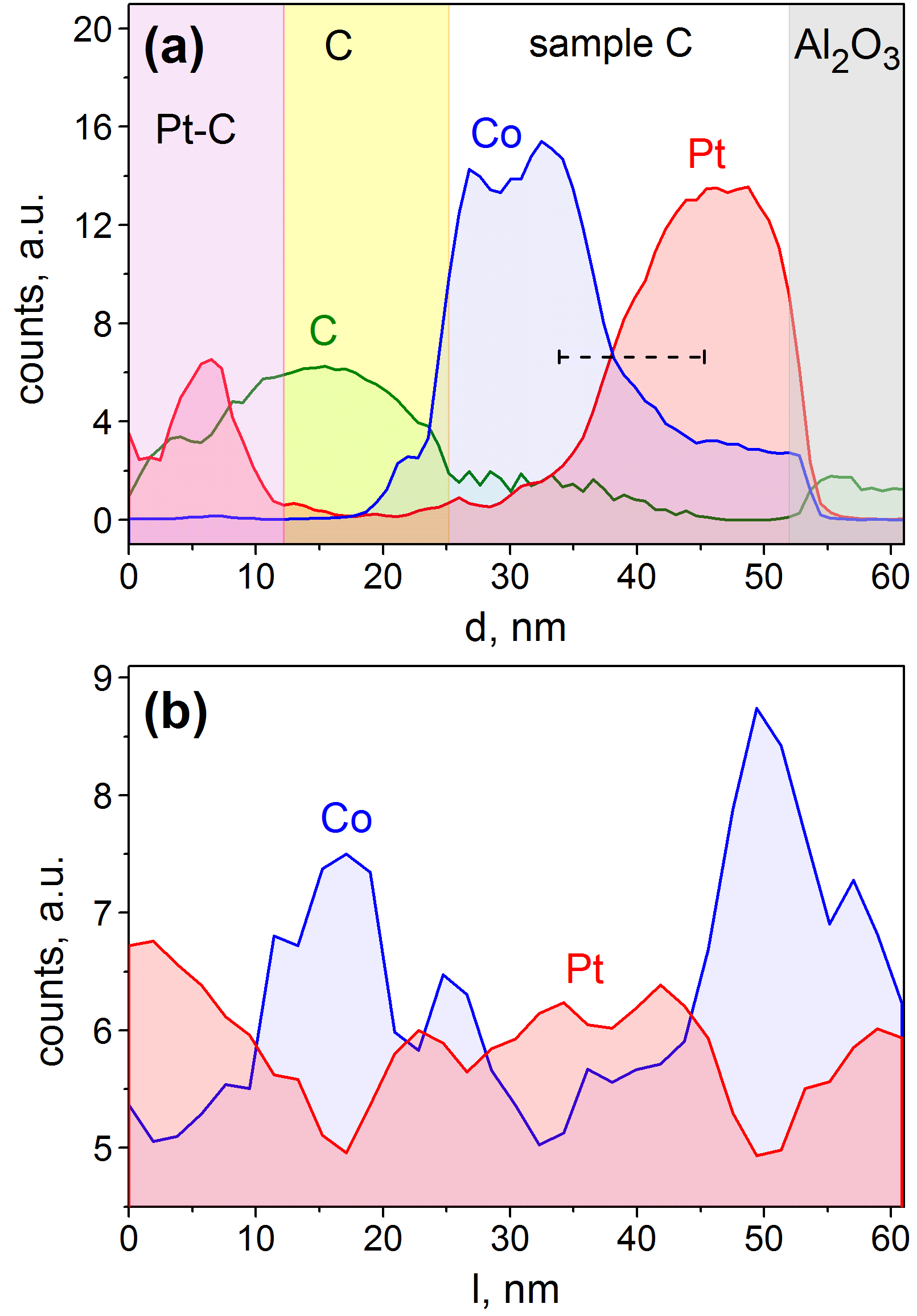}
    \caption{(a) Cross-sectional and (b) lower layer in-plane EDX spectrograms for sample C acquired along the respective arrows in Fig.~\ref{fADF}. The dashed line in (a) sketches the choice of the thickness of the control sample D where the CoPt L1$_0$ phase is expected to be formed over nearly the entire sample volume.}
   \label{fEDX}
\end{figure}

To get insight into the microstructure of the purified samples and to examine, whether the assumed CoPt L1$_0$ phase is indeed present in sample C, once its magneto-resistance measurements had been completed, we inspected sample C by STEM. Figure~\ref{fADF} presents cross-sectional TEM images of sample C in the high angle annular dark field mode (a) and in the annular dark field mode (b). The respective spectrograms obtained by STEM-EDX along the direction depicted by the arrows in Fig.~\ref{fADF}(b) are shown in Fig.~\ref{fEDX}.
We now consider the TEM and EDX data in detail. From the cross-sectional STEM-EDX spectrum in Fig.~\ref{fEDX}(a) it follows that the top layer of sample C predominantly consists of Co with a very minor content of Pt and C, whereby the Pt content gradually increases upon reaching the Co/Pt interface. The bottom layer largely consists of Pt with a notable content of Co down to the Al$_2$O$_3$ substrate, see the ``step'' in the Co signal profile in Fig.~\ref{fEDX}(a). The black region above the Co layer in Fig.~\ref{fADF}(b) is a carbon-rich layer peculiar to the TEM lamella preparation. When taking a closer look at the TEM micrograph in Fig.~\ref{fADF}(b), one recognizes a series of light channels running through the entire thickness of the bottom layer. The in-plane scan, acquired within the bottom layer and shown in Fig.~\ref{fEDX}(b), reveals that these light channels correspond to Co-rich areas in the Pt-rich layer. The substantial variation of the Co and Pt signals in the in-plane scan further corroborates the hypothesis that the pores emerged in the course of purification of the Pt layer have been filled with Co.~

\begin{figure*}
\centering
    \includegraphics[width=0.78\textwidth]{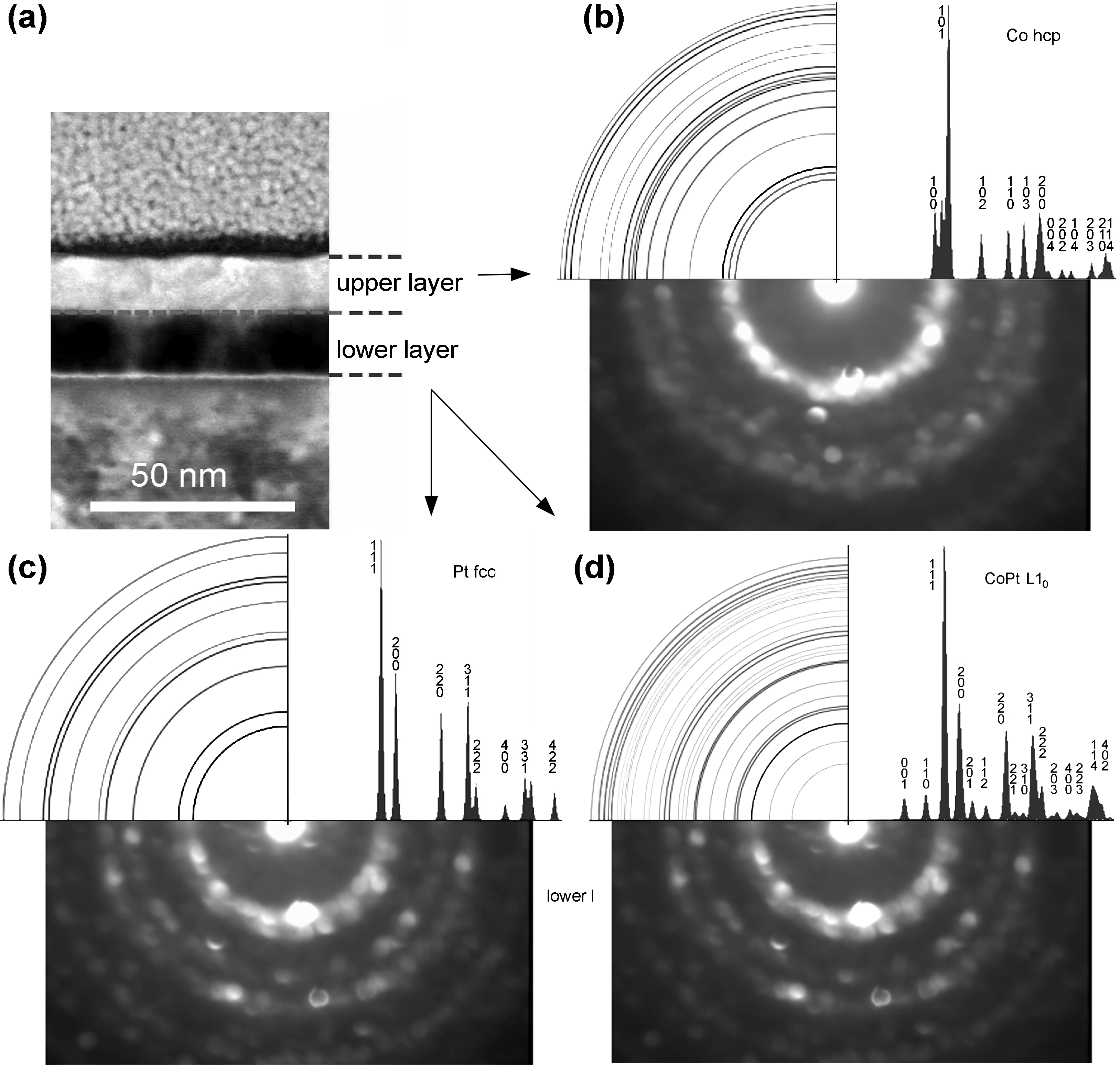}
    \caption{The location of the probed layers is shown in panel (a). Nano-diffractograms of the upper (b) and the lower (c,d) layer of sample C alongside with the simulated diffraction patterns for a Co hcp phase (b), a Pt fcc lattice (c) and a CoPt fct phase (d). }
   \label{fDIF}
\end{figure*}

The individual nanodiffraction images for the upper and the lower layer are shown in Fig.~\ref{fDIF}. The diffractograms are accompanied by the respective simulated diffraction patterns. Among the reflections in the upper layer in Fig.~\ref{fDIF}(b) one recognizes the intensive (100)+(101) rings and clearly visible (110) and (200) rings which are the fingerprint for a Co hcp lattice. The rings (102), (103), and (114) may also be recognized, though these have a much lower intensity. As for the reflections for lower layer, we compare these with a Pt fcc lattice in Fig.~\ref{fDIF}(c) and a CoPt fct phase in Fig.~\ref{fDIF}(d). As the simulation patterns depict, the bright rings (111), (200), (220) and (311) are expected for both lattices while the main reflections are dominated by Pt. At the same time, a weak additional diffraction intensity within the innermost Pt (111) ring suggests the presence of some smaller contribution from a CoPt fct phase, thereby supporting our hypotheses that the CoPt L1$_0$ phase is formed in the lower layer. For comparison, no such intensity is visible for Co in the upper layer. At the same time, we believe that no full transformation to the L1$_0$ phase took place in the lower layer, but a partial transformation on the large inner surface of the nanoporous Pt layer in which the Co deposit (and then purified Co) is located. Accordingly, the diffraction pattern of the lower layer most likely shows an overlay of the Pt and the CoPt L1$_0$ phases.~
\begin{figure}[b]
\centering
    \includegraphics[width=0.4\textwidth]{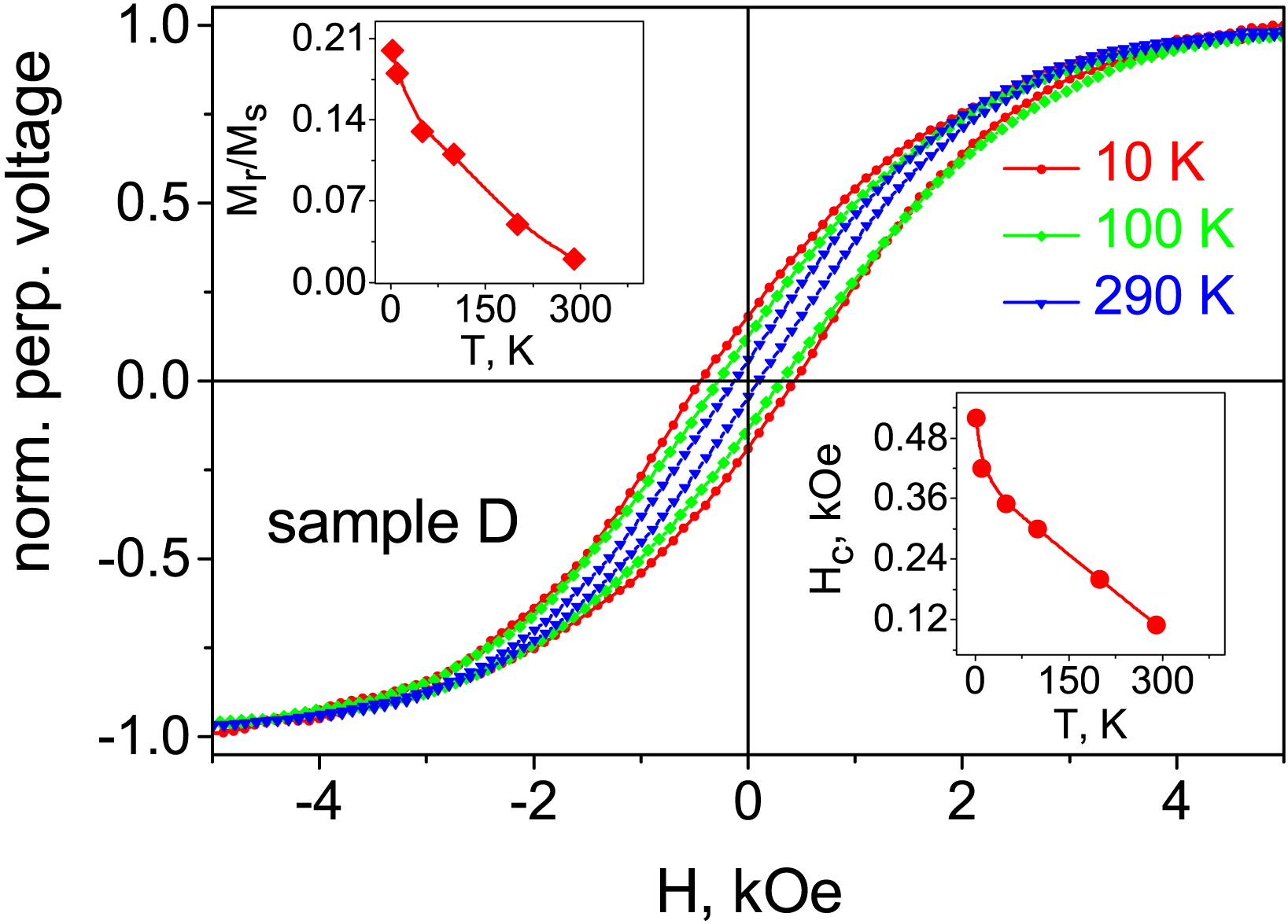}
    \caption{Isothermal Hall voltage cycling for sample D at a series of temperatures, as indicated. Insets: Temperature dependences of the squareness $M_r/M_s$ and the coercive field $H_c$ for sample D. The lines are guides for the eye.}
   \label{fFPT}
\end{figure}

\subsection{Hard-magnetic response at a Co/Pt ratio of $\mathbf{1:1}$}
As the presence of the CoPt L1$_0$ phase is confirmed by TEM inspection, we next examine the assumption that the hysteresis development and the rectangularity enhancement are indeed due to the presence of the CoPt L1$_0$ phase in the processed samples. For this reason a control sample D was prepared, with the entire thickness chosen as shown by the dashed line in Fig.~\ref{fEDX}(a). The thickness of the Co layer in sample D was chosen such that, given the nano-porosity of the processed platinum, its atomic content per volume was set to be nearly equal to that in the processed Pt layer. In consequence of this, sample D is a nano-stripe where the formation of the CoPt L1$_0$ phase is most favorable (the Co/Pt ratio is very close to $1:1$) and this phase is expected to be formed over nearly the entire sample volume. This is in contrast to samples B and C, where the CoPt L1$_0$ phase is likely formed within an interface layer only.~

The Hall voltage cycling for sample D is shown in Fig.~\ref{fHall}D and \ref{fFPT}. It demonstrates a mostly hard-magnetic behavior. The $U(H)$ curve exhibits the most open, rectangular hysteresis loop among all measured samples, with $H_c = 0.5$~T and a squareness $M_r/M_s$ of $0.18$. This provides strong evidence that magnetic response hardening in the processed CoPt-FEBID nano-stripes is indeed due to the CoPt L1$_0$ phase, that is, in turn, in agreement with the correlation between the magnetization saturation field and the Co/Pt ratio depicted in the inset to Fig.~\ref{fHall}A. Indeed, the reduction of the saturation field $H_s$ with reduction of the Co/Pt ratio can be explained by the increasing perpendicular magnetocrystalline anisotropy.
\enlargethispage{1\baselineskip}

The Hall voltage cycling $U(H)$ for sample D was repeated at different temperatures up to room temperature, see Fig.~\ref{fFPT}. The temperature-induced reduction of the coercive field and the remanent magnetization is presented in the inset to Fig.~\ref{fFPT}. A linear extrapolation of the $H_c(T)$ data suggests that above $400$~K sample D will behave as paramagnet, attesting to the robustness of the ferromagnetism in this sample at room temperature.

\section{Conclusion}
To summarize, we present an approach allowing for a controllable tuning of the magnetic properties of nano-stripe layered Co/Pt heterostructures with high resolution on the lateral mesoscale. We have demonstrated that by means of post-growth irradiation and heating of samples as well as by pre-defining the layer thicknesses, the magnetic response of the nano-stripes can be locally tuned from the soft-magnetic properties of Co to the hard ferromagnetic response of the CoPt L1$_0$ phase. The reported approach is relevant for basic research in micro-magnetism and spin-dependent transport, as well as for various applications.

\begin{acknowledgements}
HP thanks Prof. Ferdinand Hofer, Prof. Werner Grogger, Prof. Gerald Kotleitner, and Martina Dienstleder for support. HP acknowledges financial support by the EU FP7 programme (FP7/2007-2013) through grant No. 312483 (ESTEEM2). Financial support by the DFG under Grant No. HU 752/8-1 is acknowledged. This work was conducted within the framework of the COST Actions MP1201 (NanoSC) and CM1301 (CELINA) of the European Cooperation in Science and Technology.
\end{acknowledgements}

\clearpage

\vspace{7mm}
$^*$mailto: Dobrovolskiy@Physik.uni-frankfurt.de\\

April 01, 2015\\

\end{document}